\documentclass[%
 aip,
cp,  
 amsmath,amssymb,
 reprint,%
]{revtex4-2}

\usepackage{graphicx}
\usepackage{dcolumn}
\usepackage{bm}
\usepackage{graphics} 
\usepackage{subcaption}
\usepackage{mathptmx} 
\usepackage{times} 
\usepackage{amsmath} 
\usepackage{amssymb}

\usepackage[utf8]{inputenc}
\usepackage[T1]{fontenc}
\usepackage{mathptmx} 

\usepackage[utf8]{inputenc}
\usepackage[T1]{fontenc}
\usepackage{mathptmx} 

\begin{document}

\title{Driving-Over Detection in the Railway Environment}

\author{Tobias Herrmann} 
 \email{he@bahntechnik.de.}
\affiliation{
  Institut für Bahntechnik (IFB), Berlin, Germany.
}

\author{Nikolay Chenkov}
 \email{nikolay.chenkov@industrial-analytics.io.}
\author{Florian Stark}%
 \email{florian.stark@industrial-analytics.io.}
\affiliation{%
Industrial Analytics GmbH, Berlin, Germany.
}%
\author{\\Matthias H\"arter}
 \email{matthias.haerter@deutschebahn.com.}
\author{Martin K\"oppel}%
 \email[Corresponding author: ]{martin.koeppel@deutschebahn.com.}
\affiliation{%
DB InfraGO AG, Berlin, Germany.
}%

\date{\today} 

\begin{abstract}
To enable fully automated driving of trains, numerous new technological components must be introduced into the railway system. Tasks that are nowadays carried out by the operating stuff, need to be taken over by automatic systems. Therefore, equipment for automatic train operation and observing the environment is needed. Here, an important task is the detection of collisions, including both (1) collisions with the front of the train as well as (2) collisions with the wheel, corresponding to an  driving-over event. Technologies for detecting the driving-over events are barely investigated nowadays. Therefore, detailed driving-over experiments were performed to gather knowledge for fully automated rail operations, using a variety of objects made from steel, wood, stone and bones. Based on the captured test data, three methods were developed to detect driving-over events automatically. The first method is based on convolutional neural networks and the other two methods are classical threshold-based approaches. The neural network based approach provides an mean accuracy of 99.6\% while the classical approaches show 85\% and 88.6\%, respectively. 
\end{abstract}

\maketitle
\section{INTRODUCTION}

Less road traffic, fewer traffic jams, less environmental pollution - and more people and more goods on the rails. The rail sector in Europe is on the verge of a technological leap into the digital future. The "Digital Rail Germany" sector initiative is seizing this opportunity and bringing future technologies to the rail system. The foundation for this is being laid with the fundamental modernization and digitization of the infrastructure through the consistent introduction of digital control and safety technology. 

Several technological challenges are associated with the goal of having trains on mainline railroads run in a highly automated manner in the future. 
In automatic train operation (ATO) \cite{gebauer2012autonomously}, technical systems take over tasks that had previously been performed by the operating staff. ATO includes different Grades of Automation (GoA)(i.e. Tab. \ref{goas}), up to GoA4 in which the train is fully automated with no staff on board. 
One of the challenges is to detect possible collisions and driving-over objects events and to derive appropriate responses for the driving operations. The variety of objects in the track area is manifold and ranges from natural objects to metal parts and living beings.

Based on the expected events in regular rail operations, initial requirements for an impact and driving-over detection system were formulated. In order to be able to detail the requirements, a feasibility study was carried out. Based on the findings of the feasibility study, test runs were planned to (1) record driving-over events and (2) develop approaches for detecting driving-over events. The following article outlines the driving-over experiments and proposes three methods to detect such events automatically.

\section{STATE-OF-THE-ART}

Fully automated rail operations \cite{overview} have been used so far primarily in closed (metro) systems and are not yet state-of-the-art in mainline rail operations. Closed systems have no interfaces to the environment (forest adjacent to the tracks, crossing roads, etc.) and therefore other safety requirements. 
A comprehensive overview of various projects on digitization and automation in rail operations in Germany is provided by the 2019 EBA research report \cite{EBA}. It summarizes projects from the areas of digital driver assistance systems, digitization, and automation. At the international level, the "World Report on Metro Automation" \cite{WorldReportOnMetroAutomation} from the International Association of Public Transport (UITP) provides a comprehensive global overview of automated rail operations (metros). Automatically operated metro systems are in operation in 15 cities \cite{AllianzProSchiene}. Globally, a total of 64 fully automated metro lines were operating in 42 cities as of March 2018 \cite{UITP}. The grade of automation (GoA) in this context has been defined by UITP \cite{UITP},
 as shown in Tab. \ref{goas}.
Internationally, the fully automated train operation of the mining company Rio Tinto received much attention [5], [6].  A 2.4 km long freight trains in Australia was fully automated, driving without a train driver.  This system is of particular interest because it is a retrofitted system that operates on the regular network and is not a closed metro system. The level of automation here can be given as GoA 4. In Germany, the Skytrain (people mover) and the U2 and U3 subway lines of the Nuremberg subway system [7], which have been in operation for several years, are particularly worth of mention with regard to automated traffic. Since the topic of fully automated rail vehicles is becoming more and more important, there are several industrial and research projects in which automated operation according to GoA 2 to GoA 4 or the way to it is being tested \cite{Spiegel}.

All of these systems have the common feature that objects and obstacles in the path are detected via various camera, radar and/or lidar systems and, on the basis of this object detection, emergency braking is or could be initiated to stop the vehicle \cite{assaf2022high}, \cite{zhangyu2021camera} \cite{petrovic2022integration}. A disadvantage of this setup with camera, radar and/or lidar systems is that it is not known whether a collision has actually occurred between the vehicle and the object in the form of an impact on the front structure, an driving-over event or a combination of both. This requires sensors that can detect the direct collision with positional accuracy and, optionally, possible damage. First collision detection systems have been introduced in \cite{RioTinto_2}, \cite{RioTinto_1} and in the subway system Nuremberg. 
In \cite{RioTinto_2}, \cite{RioTinto_1} an acceleration sensor is mounted on the front of the train. This sensor is able to detect collisions on the train front. Due to the limited sensor setup, it is difficult to estimate damage to the train. Moreover, driving-over events can hardly be detected. An human interaction is also required when an event have been detected by the system. Then, images are transmitted to a control station, where a person evaluates the collision.
The subway in Nuremberg uses a mechanical device to detect driving-over events \cite{PatentSiemens}, \cite{Siemens2}. A bracket is installed on the train, just above the track. In the event of a collision, an object on the track causes this bracket to fold back mechanically, thus signaling a collision. The disadvantage of this system is that it is only possible to determine in binary terms whether an event has occurred or not.

In case of an collision or driving-over events, responsible authorities have to be informed and the operational procedure may have to be changed significantly (e.g. immediate stop of the vehicle). Therefore, detection systems for such events are absolutely necessary in a fully automated railroad system with interfaces to the environment.

Currently, there are no datasets available for the detection of collisions in the rail vehicle sector. Therefore, this project was initialized to test technologies for detecting driving-over events. The results from the project are presented in the following sections.

\begin{table}
  \centering
  \vspace*{+0.25cm}
  \caption{Simplified description of GoA from DIN EN 62290-1.}
  \begin{tabular}{|l || c |}
    \hline
    \textbf{GoA}  & \textbf{Train operation}  \\
    \hline
    0 & On-sight with no automation\\ 
    \hline
    1 & Non-automated according to signals\\
      & and with safeguarding in case of overlooked signals      \\ 
    \hline
    2 & Semi-automatic with acceleration and deceleration \\ 
      & controlled by a technical system \\ 
    \hline
    3 & Driver-less with a train attendant    \\ 
    \hline
    4 & Unattended without crew on board  \\ 
    \hline
  \end{tabular}
  \label{goas}
\end{table}

\section{OVERVIEW}
In this publication a prototypical system for detecting driving-over events is described. To the best knowledge of the authors, the proposed approach is the first non-mechanically operating driving-over detection method system for railway vehicles.

In Sec. \ref{sec:testsetup}, the test setup and the test procedure are outlined in  and Sec. \ref{sec:testprocedure}. For the tests a freight wagon was equipped with 42 sensors using four different measuring modalities. The wagon drove with three velocities, i.e. 5 km/h, 10 km/h and 15 km/h, over objects types that can appear in regular operation. Based on the captured signals, three detection approaches have been developed (cf. Sec. \ref{sec:detectionmethods}). One is based on Convolutional Neural Network (CNN) and the others are classical threshold methods. Finally, the results are described in Sec. \ref{sec:experiments}

\section{TEST SETUP} \label{sec:testsetup}
In this section the test setup is outlined. First, the general setup is explained and then measurment setup is outlined

\subsection{General Test Setup}

The variety of objects that can be crossed by rail vehicles is manifold and ranges from living beings of different sizes to natural objects (e.g. fallen trees) and man-made objects (e.g. forgotten construction equipment, roof tiles). To account for this diversity,
the possible object classes must be abstracted. Thus, the drive-over tests can took place within a controlled experimental framework. Animal bones of various sizes, roof tiles, birch sticks (approximately 50 mm in diameter), laminated wood (20 mm x 55 mm), and various steel elements (flat steel and steel wedge 6 mm to 30 mm in height) were identified and procured or fabricated as possible abstractions (cf. Fig. \ref{fig:objects}.

Another important factor was the selection of the test vehicle. Its selection was based on the decision that it should first be investigated in principle, whether detection of driving-over events is possible. Hence, a type Res freight wagon (unladen mass: 24.5 t), equipped with type Y25 bogies, was selected as the test vehicle and fitted with sensor technology. Reasons for selecting this vehicle for the test operation include its good availability, easy transport to the test track, and limited financial damage in the event of a possible accident in the controlled test environment.

\subsection{Measurement Test Setup}

Different sensor types were considered for the detection of the driving-over events. During the tests, the use of the following sensor types was considered:

\begin{itemize}
\item uniaxial and triaxial capacitive accelerometers (with different measurement ranges) on different components of the wagon,
\item microphones (including wind screen) each positioned close to and aligned with the direction of the leading wheel set,
\item laser distance sensors directed from the trolley towards the rail.
\item wire draw, i.e. displacement, sensors at several positions at the wagon.
\end{itemize}

\begin{figure}[t]
      \vspace*{+0.25cm}
      \includegraphics[width=\linewidth]{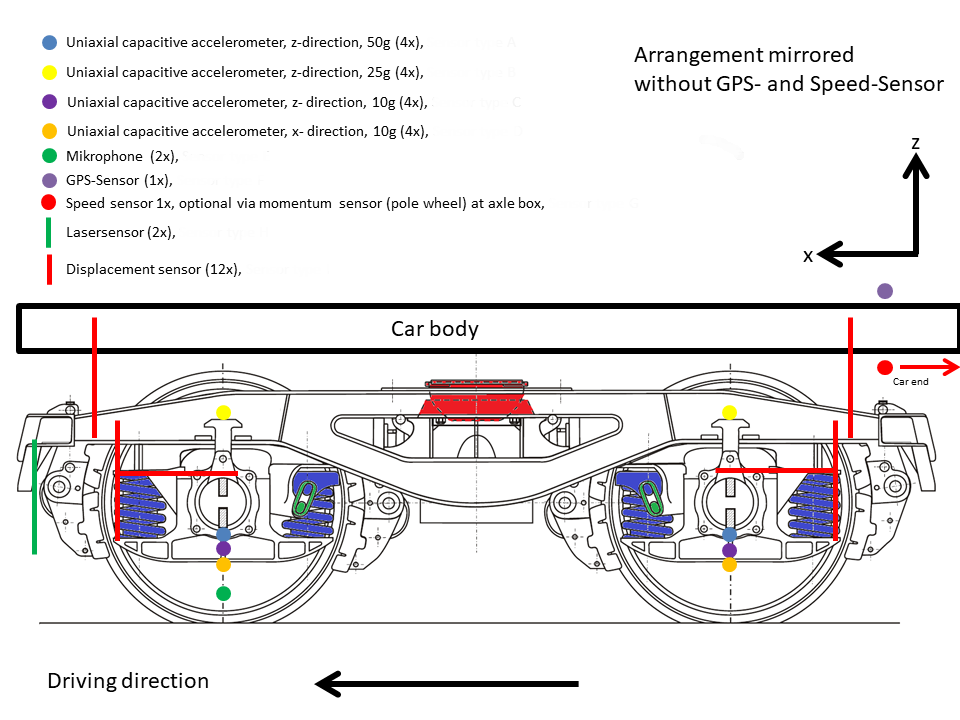}
      \caption{Left side view of the leading bogie with applied sensors.}
      \label{FL-1}
\end{figure}

\begin{figure}[t]
      \vspace*{+0.25cm}
      \includegraphics[width=\linewidth]{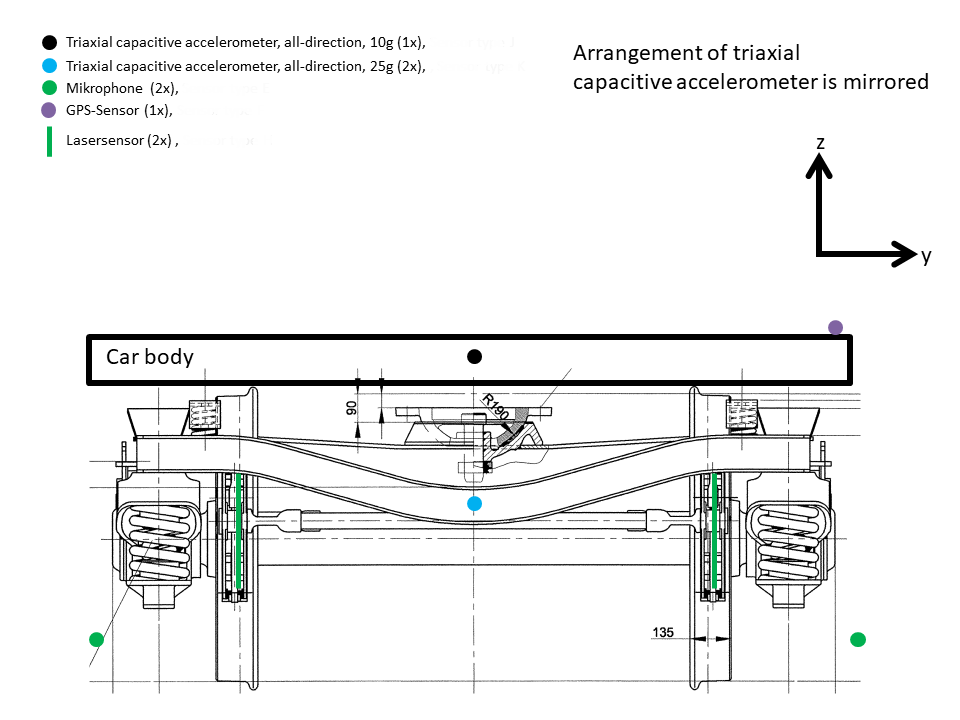}
      \caption{Front view of the leading bogie with applied sensors.}
      \label{FL-2}
\end{figure}

The first two types of sensors detect the result of the driving-over event event, while the laser distance sensor should detect the object on the rail a few moments before the driving-over. Capacitive accelerometers were used to calculate the wheelset lift as a result of passing over objects, which allow static accelerations (e.g., gravity) to be detected. As a consequence of the change of the wheel-rail contact to a wheel-object-rail contact, noise emissions are to be expected. These can be recorded with the microphones. The laser distance sensors enable measurement of the distance between the carriage and the rail and can thus detect objects located on the rail.

\begin{figure}[t]
      \vspace*{+0.25cm}
      \includegraphics[width=\linewidth]{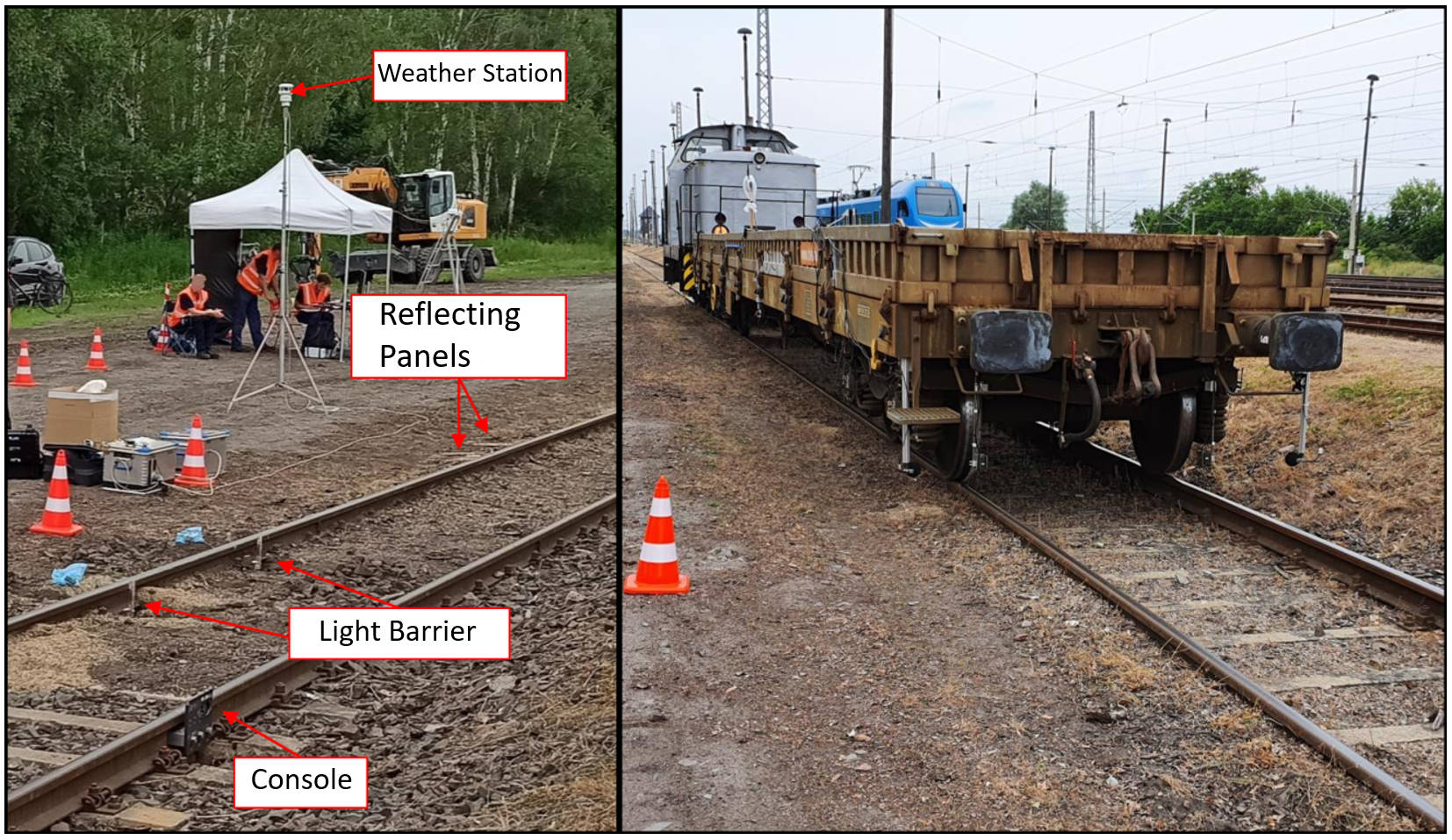}
      \caption{Test setup on the track side.}
      \label{FL-3}
\end{figure}

For the validation of possible later simulations, draw wire, i.e. displacement, sensors were attached at various positions on the wagon. These were primarily not considered for driving-over detection due to their inertia. A GNSS sensor and an encoder were used for position and speed measurement. Furthermore, a laser light barrier was attached to the wagon body, which enabled precise local synchronization with reference to the object to be crossed with the aid of reflective train closure panels positioned in the track area (Fig. \ref{FL-3}).

Fig. \ref{FL-1} and \ref{FL-2} show the sensor types and their positions on the leading bogie schematically. The sensors were applied on the right side analogous to the left side. Both the microphones and the laser distance sensors were installed in front of the leading bogie using brackets. The acceleration sensors were positioned as follows: 

\begin{itemize}
\item at the wheelset bearing,
\item on the bogie frame directly above the primary spring step,
\item centrally on the front and rear (not shown) endcarriages of the bogie frame,
\item on the wagon body in the center of the front-loading area in the direction of travel.
\end{itemize}

Two sensors each with different measuring ranges or sensitivities in the vertical direction and one sensor each in the longitudinal direction were used on the wheelset bearing. The vertical and longitudinal directions are the directions in which the strongest reactions are to be expected as a result of an driving-over event. The different measurement ranges were chosen to determine events with different resulting acceleration amplitudes as accurately as possible. The relevant frequency range of the acceleration signals was estimated to be 0 Hz to 1 kHz. For this purpose, the sampling frequency was set to 5 kHz and a low-pass filter with a cutoff frequency of 1 kHz was applied. The microphone signals were captured with a sampling frequency of 50 kHz and an anti-aliasing filter with a cutoff frequency of 20 kHz.

In addition to the freight car, the track was also equipped with sensors (see Fig. \ref{FL-3}). A weather station recorded the wind speeds, the air temperature, the humidity and the air pressure. The recording was triggered when the first wheel-set passed through the installed light barriers a few moments before hitting the object to be passed over. These objects were mounted in a specially designed and built console to prevent the test objects from rolling away or falling down (cf. Fig. \ref{fig:objects}). A total of two light barriers were positioned at a defined distance from each other, from the console and from the reflective train closure panels. This allowed local and temporal synchronization between the vehicle and the obstacle, and the current velocity could be determined for each passing wheel-set. The temporal synchronization between the measurement system on the track and that on the vehicle was achieved via the GPS time signal.
\section{TEST PROCEDURE}\label{sec:testprocedure}

The tests were carried out on four consecutive days. Beforehand, both the track and the vehicle were prepared. On the first day of the trials, comparative runs at different speeds were carried out in the marshalling yard. By means of these comparative runs at different speeds over several switches, measurement data for regular operation could be generated. These measured data is used for comparison against the measured data from the driving-over events. In this way, it can be checked whether the driving-over events can be distinguished from signals appearing in regular rail operations caused by switches, rail faults, etc. 

Before starting the test runs with objects, test runs were performed at different speeds and without passing objects on the prepared track. For each of the tested objects, two test runs were then performed at speeds of 5 km/h, 10 km/h and 15 km/h respectively. The test vehicle was first accelerated to the target speed by a pushing shunting locomotive and then pushed off. The test vehicle then rolled over the test section with the attached objects. After passing through the test section, the wagon was stopped with braking shoes. The complete process was recorded via the sensors and additionally installed cameras. While Fig. \ref{FL-1}, \ref{FL-1} and \ref{FL-1} show an overview of the general test setup on the infrastructure and the test vehicle, Fig. \ref{fig:objects} shows examples of how the objects were positioned on the rail using a console constructed for the tests.

\section{DETECTION METHODS}\label{sec:detectionmethods}

Three methods where developed, for detecting driving-over events. One method is based on deep learning and the other two methods are classical threshold approaches. In the following, the methods are outlined in more detail. 
\subsection{Deep Learning Approach}

The first method is based on Convolutional Neural Network (CNN) \cite{KIRANYAZ2021107398}. The concept of convolution is commonly used in machine learning for data preprocessing, but also in neural networks models, i.e. CNNs. The advantage of CNNs is that temporal or spatial relations between neighbouring points in training data can be captured and learned by a kernel. In modeling, multiple kernels are normally applied in parallel to the same signal input, and each kernel could be fit to represent a statistical feature of the training dataset. The computed signal then shows high values when (a part of) the signal matches the kernel, zero when signal and kernel are uncorrelated, and low negative value when the kernel and the signal are anti-correlated.
\begin{figure}
     \vspace*{+0.25cm}
      \includegraphics[width=0.8\linewidth]{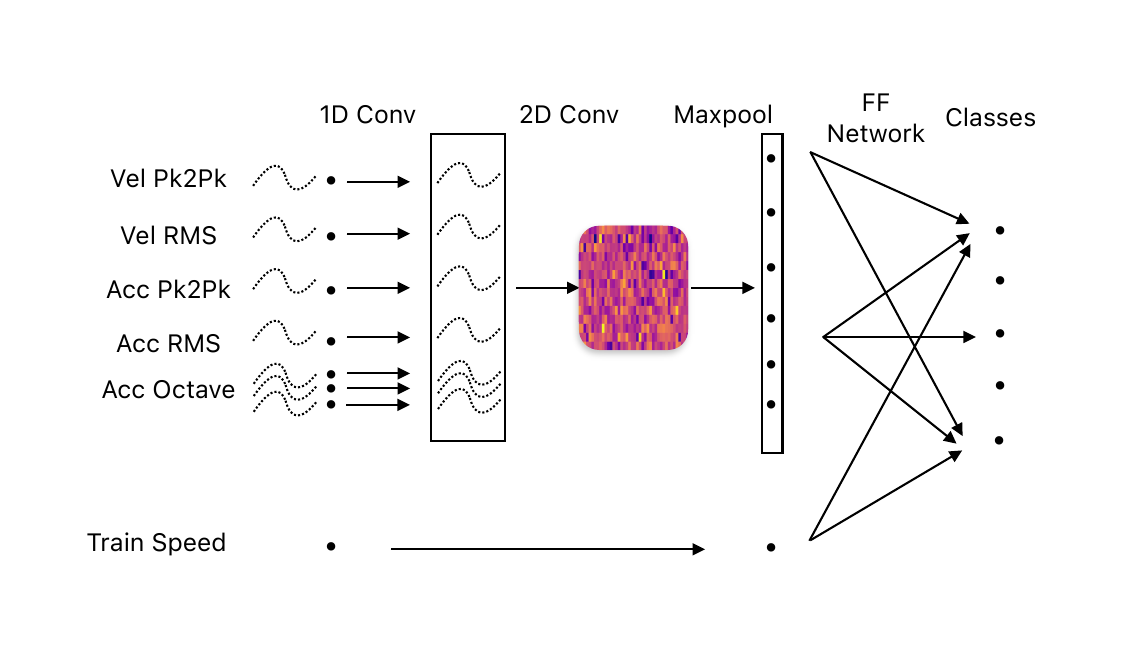}
      \caption{Convolutional network as classificator.}
      \label{DL-1}
\end{figure}
\begin{figure*}[t]
  \vspace*{+0.25cm}
  \centering
  \begin{subfigure}{0.23\linewidth}
    \centering
    \includegraphics[width=\linewidth]{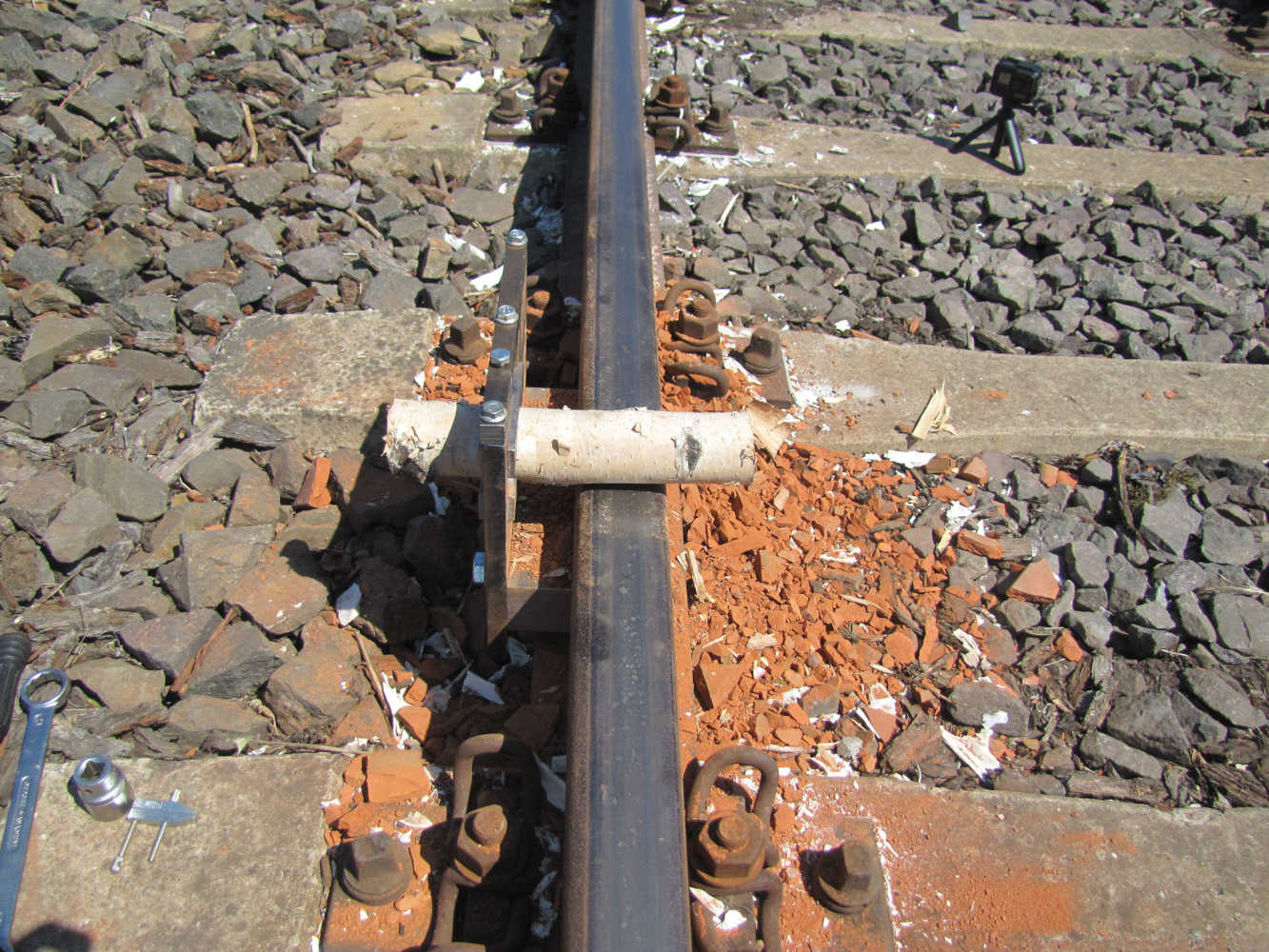}
    \caption{Birch stick}
    \label{fig:sub1}
  \end{subfigure}
  \begin{subfigure}{0.23\linewidth}
    \centering
    \includegraphics[width=\linewidth]{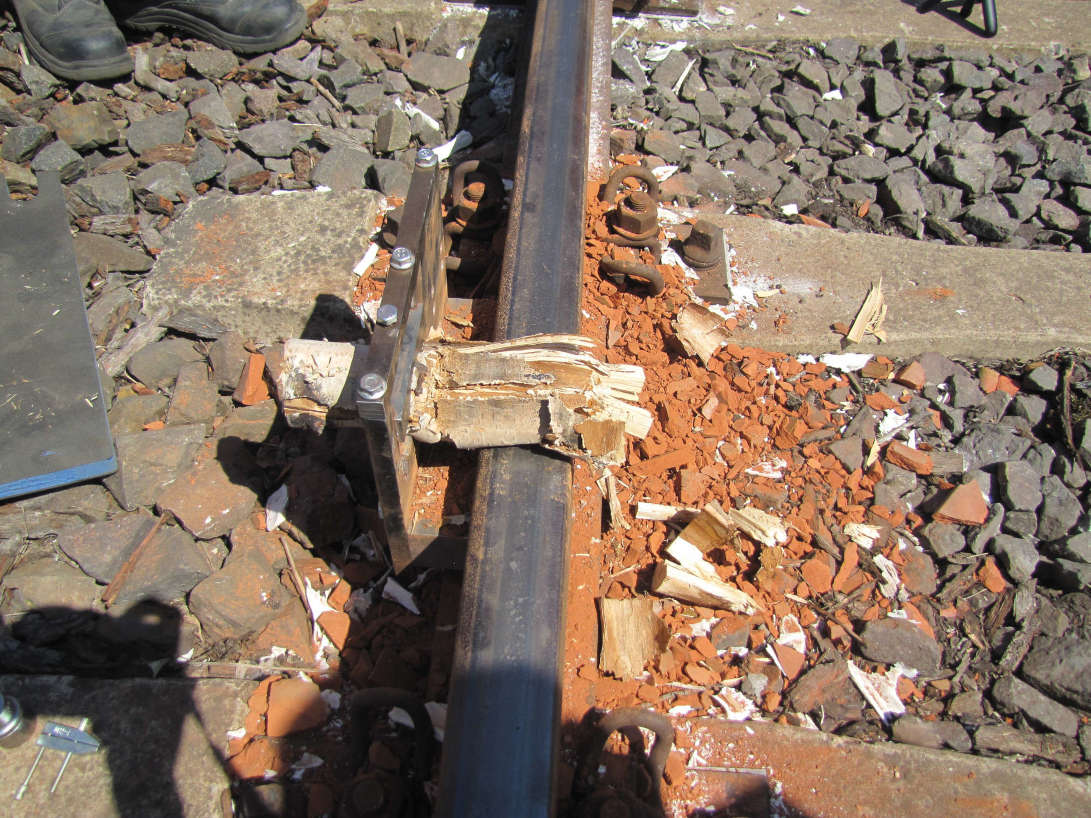}
    \caption{Birch stick}
    \label{fig:sub2}
  \end{subfigure}
  \begin{subfigure}{0.23\linewidth}
    \centering
    \includegraphics[width=\linewidth]{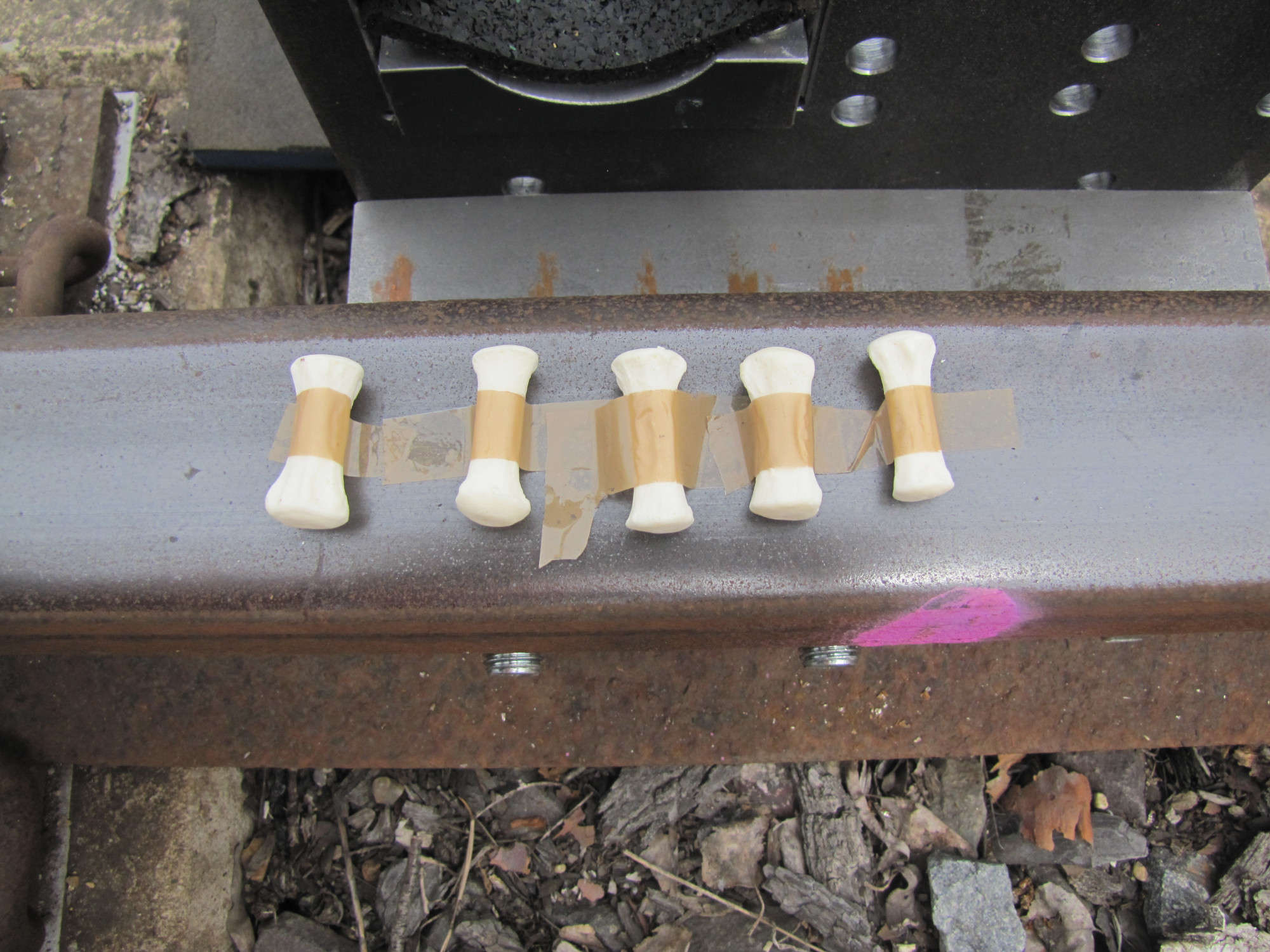}
    \caption{Small animal bones}
    \label{fig:sub3}
  \end{subfigure}
  \begin{subfigure}{0.23\linewidth}
    \centering
    \includegraphics[width=\linewidth]{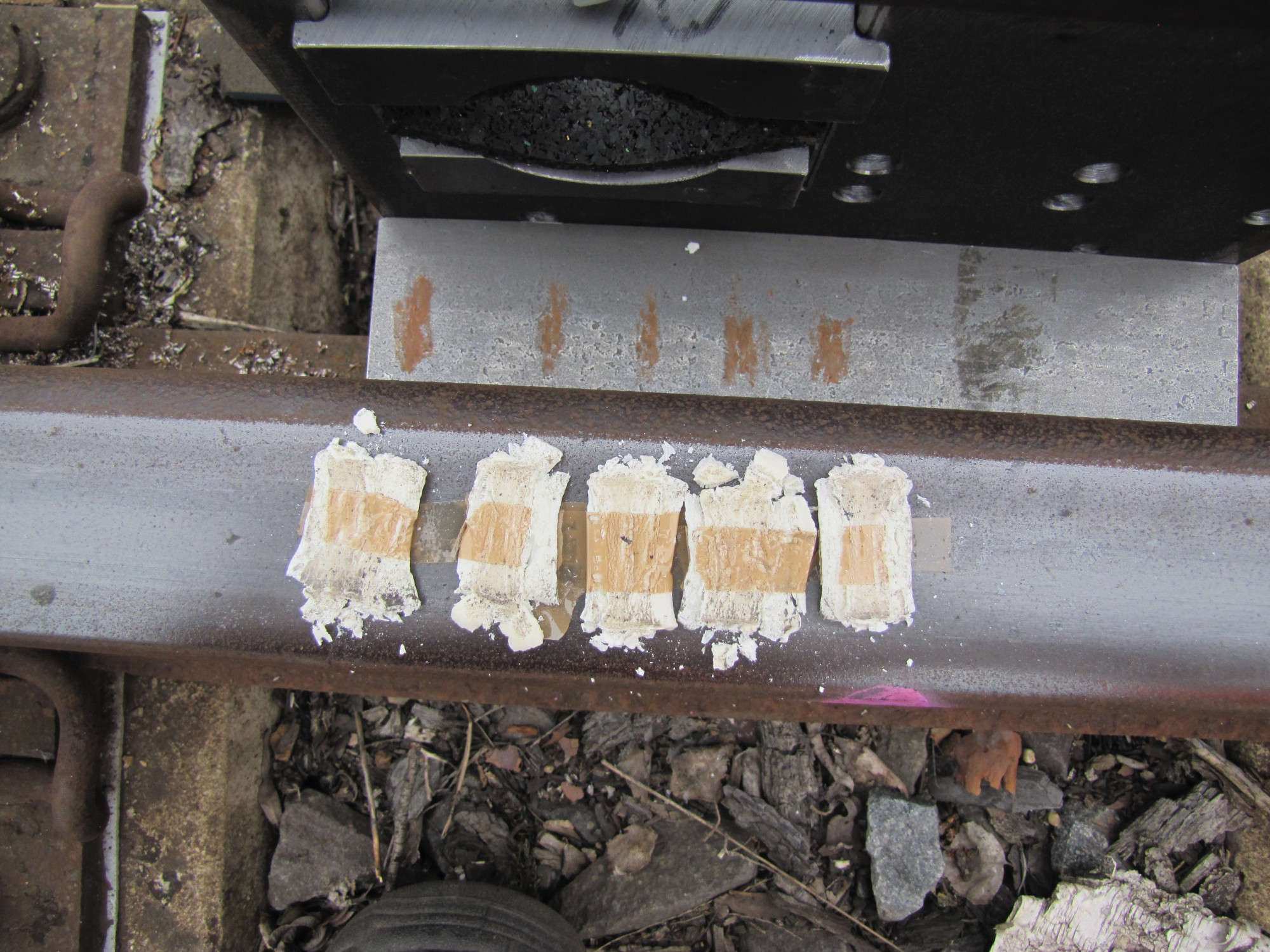}
    \caption{Small animal bones}
    \label{fig:sub4}
  \end{subfigure}
  \begin{subfigure}{0.23\linewidth}
    \centering
    \includegraphics[width=\linewidth]{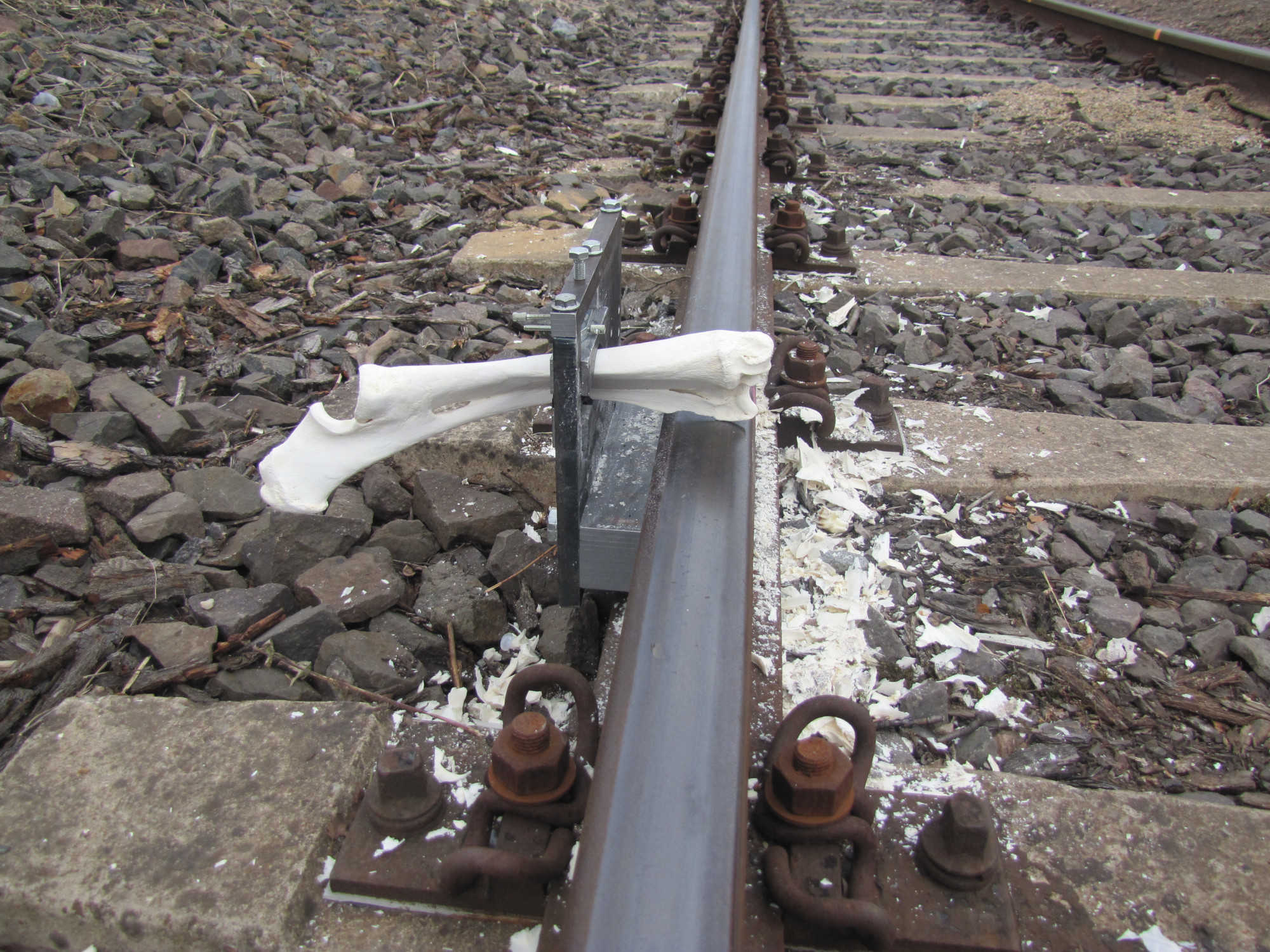}
    \caption{Large animal bones}
    \label{fig:sub5}
  \end{subfigure}
  \begin{subfigure}{0.23\linewidth}
    \centering
    \includegraphics[width=\linewidth]{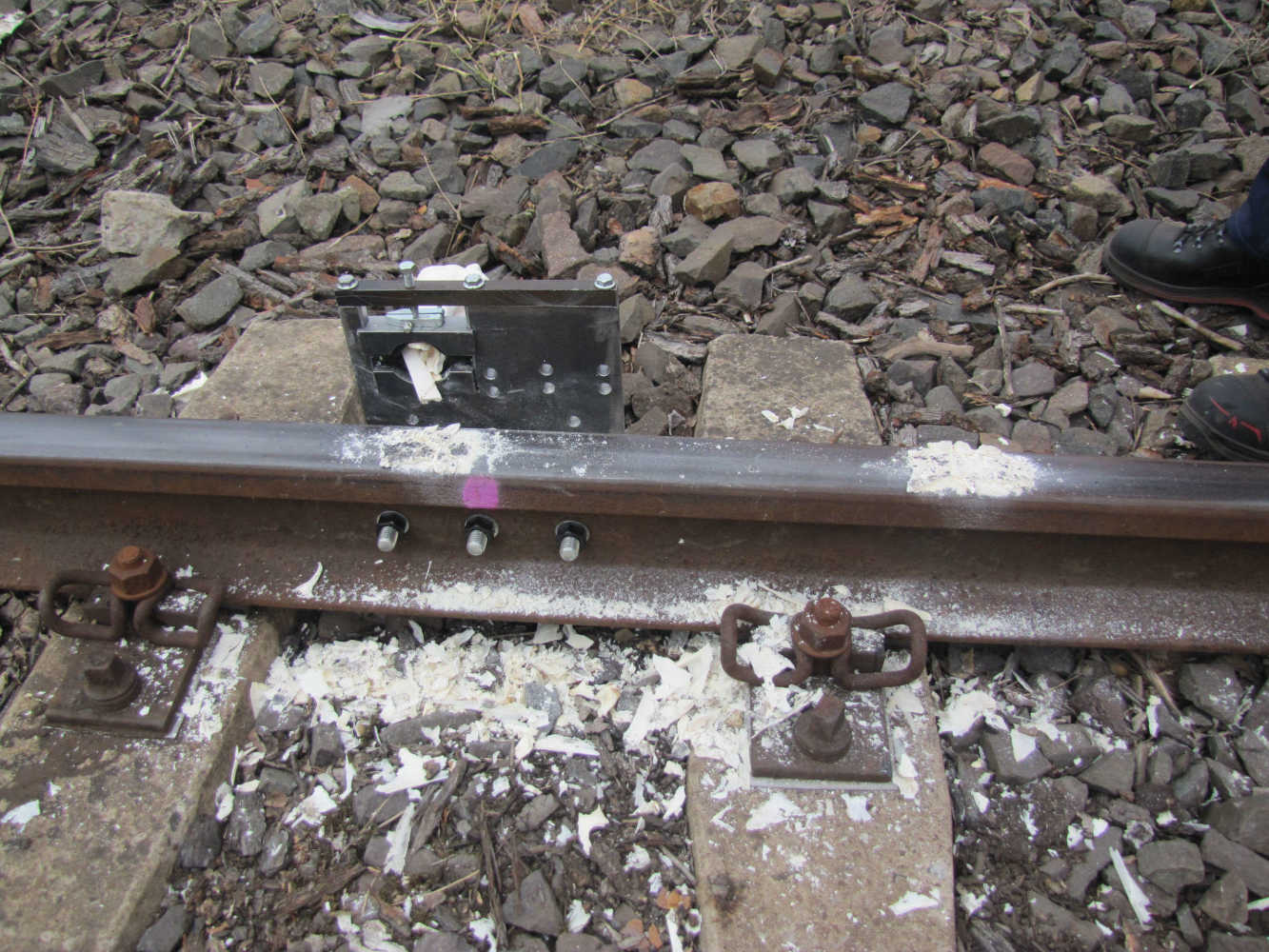}
    \caption{Large animal bones}
    \label{fig:sub6}
  \end{subfigure}
  \begin{subfigure}{0.23\linewidth}
    \centering
    \includegraphics[width=\linewidth]{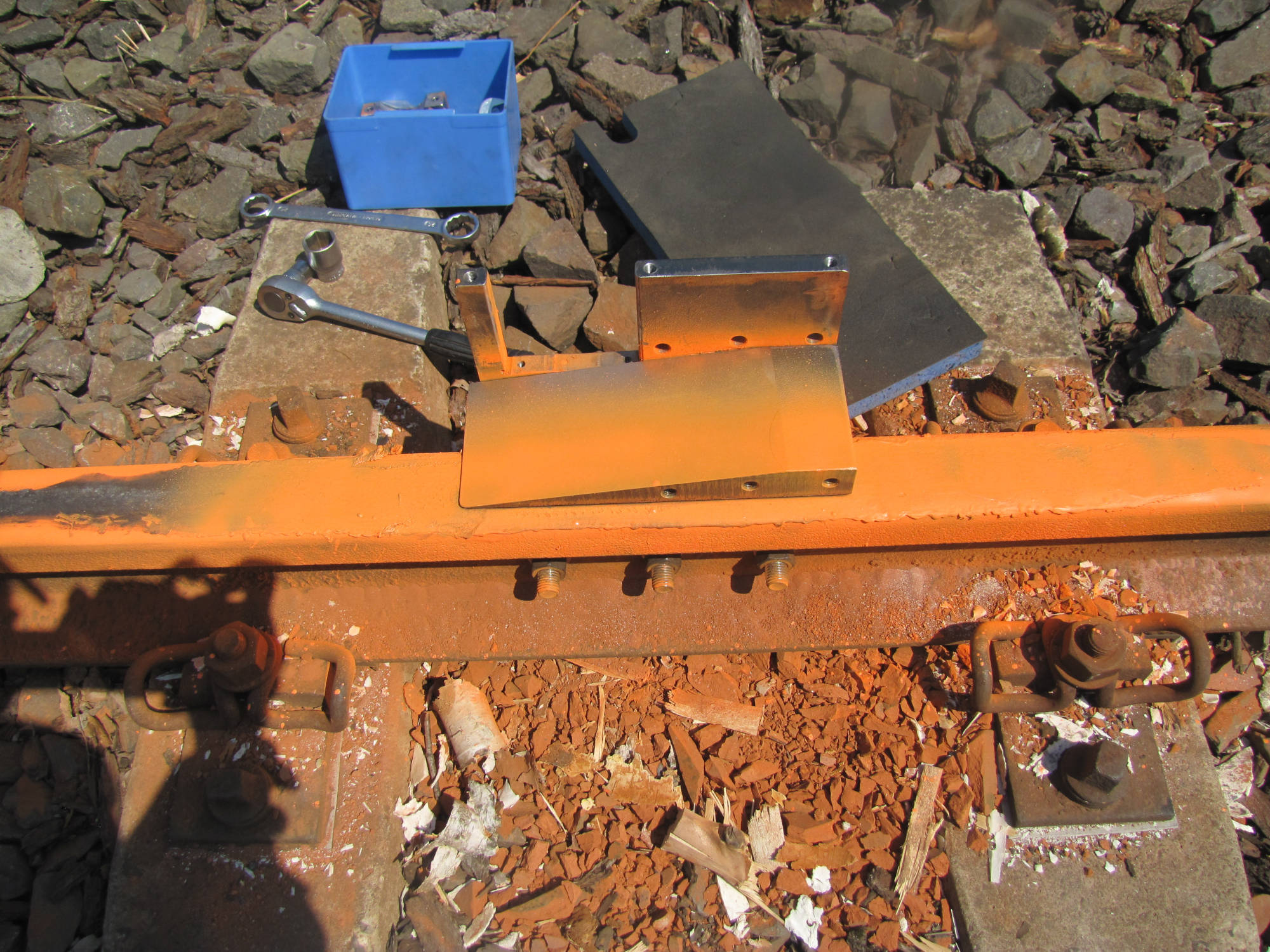}
    \caption{Steel wedge}
    \label{fig:sub7}
  \end{subfigure}
  \begin{subfigure}{0.23\linewidth}
    \centering
    \includegraphics[width=\linewidth]{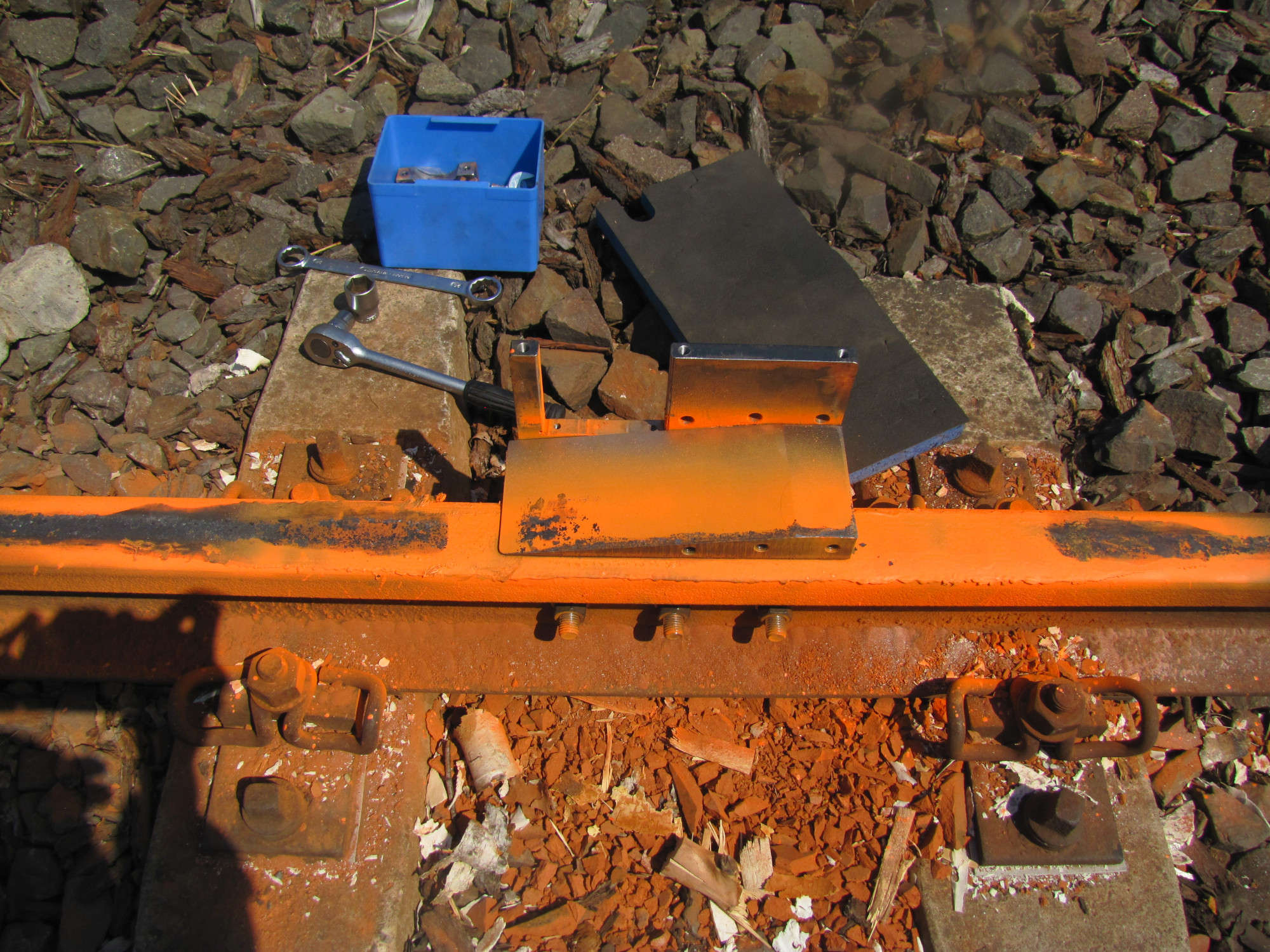}
    \caption{Steel wedge}
    \label{fig:objects}
  \end{subfigure}
  \caption{Subset of the objects used for the driving-over experiments. Fig. (a), (c), (e), (g) show the objects before the driving-over event. Fig. (b), (d), (f), (h) show the objects after the driving-over event.}
  \label{fig:objects}
\end{figure*}
\begin{figure*}[t]
   \vspace*{+0.0cm} 
   \centering
  \begin{subfigure}{0.46\linewidth}
    \centering
    \includegraphics[width=\linewidth]{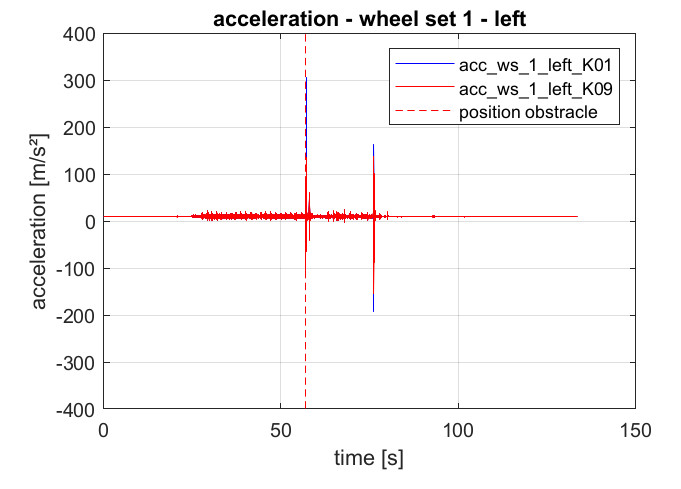}
    \caption{Birch stick}
    \label{fig:sub1}
\end{subfigure}
\hspace*{+0.25cm} 
  \begin{subfigure}{0.46\linewidth}
    \centering
    \includegraphics[width=\linewidth]{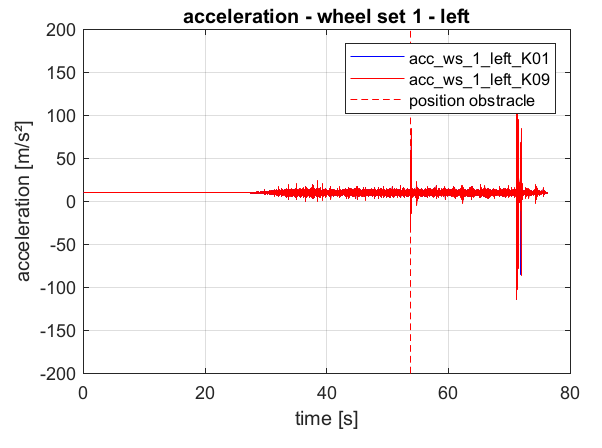}
    \caption{Animal bones}
    \label{fig:sub2}
  \end{subfigure}
  \caption{Acceleration signals in z-direction captured from sensors mounted on the wheel-set bearings at a velocity of 5 km/h. (a) Signal caused by the driving-over event of a birch stick. (b) Signal caused by the driving-over event of small animal bones.}
  \label{fig:signals}
\end{figure*}

The proposed network (dl-1) as outlined in Fig \ref{DL-1} is applied to the signals of each accelerometer. First, the high-frequency vibration data (100kHz) is preprocessed to a lower temporal resolution of 10Hz. The following quantities are computed based on the raw acceleration signal: 
(1) Velocity peak to peak, (2) Velocity root mean square, (3) Acceleration peak to peak, (4) Acceleration root mean square and (5) Acceleration octave spectrum (10 frequency bands between 100Hz and 1200Hz), one value per 100ms. These signals are then used as inputs to the network model. First, each time series is convolved with 3 kernels of width 11. The outputs of the first layer are then forwarded to an aggregator two-dimensional convolutional layer and convolved with 7 kernels with size 11x11. In the next step, a maxpool layer effectively compresses the time-series vector into a single point vector by taking the largest value in the time dimension. As the train speed is relatively constant, it's value bypasses the convolutional layers and is piped as direct input to the last stage: a feedforward network. The feedforward network consist of 2 hidden layers with SELU activation functions, except in the last layer, where softmax function is used. The number of outputs is five, equal the number of classes to be classified. The loss function is based on the cross-entropy calculation between the model output and the actual object using on-hot encoding.

\subsection{Classical Approach}
In this section the classical approaches (classical-1 and classical-vel) are outlined in more detail. The pre-processing steps are similar to the deep learning approach.
For each acceleration sensor, multiple outputs are calculated within a 100ms window, i.e. (1) acceleration peak to peak (2) acceleration root mean square (3) velocity peak to peak (4) velocity root mean square and (5) acceleration octave spectrum.
Then a threshold classifier for each of these five quantities is fit to minimize the classification error between events and non-events. To minimize the number of false positives during training, an additional weight is added to the loss function for each non-event classified as event. The threshold values are selected during a training phase, using an optimization method, i.e., simulated annealing \cite{kirkpatrick1983optimization}.
If a measured value in a time interval is greater than the predefined threshold, then the event is considered as a drive-over event. This is done for each of the five quantities. Finally, a majority voting over all quantities is carried out. Hence, if most of the quantifier signal an driving-over event, then a driving-over event is detected. 

The second classical (classical-vel) approach also takes the velocity of the train into consideration. Here, the quantities calculated from the input signal are normalized by dividing the measured values by the train velocity. The subsequent process is similar to the first classical approach.

\section{EXPERIMENTAL RESULTS}\label{sec:experiments}

During the test procedure the wagon drove over 60 objects. Fig. \ref{fig:objects} show some of the objects before and after the test. 

In a first step the sensors outcomes were manually evaluated regarding the signal plausibility. Based on these evaluations, some sensor modalities have been excluded:
\begin{enumerate}
\item The microphones show in the time domain a similar behavior as the accelerometers. Nevertheless, in some cases high values in some single frequencies could be seen and the influence of other external impacts (trains on another track, motor sound etc.) were visible and disturbed the measurement or lead to false positive. Due to the positive results of other sensors, it was decided, to exclude the microphones.
\item Due to the inertia behavior of the draw wire sensors they show non distinctive or relevant pattern. Therefore, these sensors were excluded from the test.
\item The laser sensors at the car body, that were directed to the rail surface were to prone to disturbances. In various cases the sensors show higher amplitudes for no event than for driving-over events.
\end{enumerate}

Hence, the evaluation concentrated on the accelerations sensors. For this purpose, the accelerations of the reference runs were first evaluated and the events were assigned to the known track events (switches, tracks, etc.) to the extent required for the evaluation. These measurement signals are referred to as accelerations in regular rail operations. In this way, a reference data set has been created which contains the maximum accelerations to be expected in regular operation.

The aim of the developed algorithms is to differentiate driving-over events from events during regular operation. Here, the switches leaded to the highest acceleration values. Therefore, the driving-over events were tested against data from regular operation. During the test 96 events were utilized for regular operations. 
The training was done on a random sample on 40\% of the data. We have used the same train/validation/test split 40\%/30\%/30\% for all models.

The accuracy of the models was calculated for each acceleration sensor individually. In Tab. \ref{tab:accuracy} the highest, lowest and mean accuracy's from all acceleration sensors are depicted. The lowest accuracy was measured with the sensors mounted on the car body. This is due to the fact that the vehicle body is additionally damped and events appear with lower intensity. 
On the other hand, the highest accuracy could be reached with the sensors directly mounted at the wheel-set bearings, where no damping appears. Fig \ref{fig:signals} shows signals from the acceleration sensors mounted on the wheel-set bearings. The vehicle had a velocity of approximately 5 km/h.  Fig \ref{fig:signals} (a) shows the results for the birch stick and Fig \ref{fig:signals} (a) shows the results from the small animal bones. The objects on the track caused strong acceleration signals in z-direction, that can be clearly differentiated from normal operation scenarios. 

\begin{table}[t]
  \centering
  \vspace*{+0.25cm}
  \caption{Detection Results}
  \begin{tabular}{|l || c || c || c |}
    \hline
    \hline
     & \textbf{dl-1} & \textbf{classical-1} & \textbf{classical-vel} \\
    \hline
    \textbf{Max Accuracy} & 100\% & 92.3\%   & 97.2\% \\ 
    \hline
    \textbf{Min Accuracy} & 98.7\% & 77.5\%  & 77.6\% \\
    \hline
    \textbf{Mean Accuracy} & 99.6\% & 85.3\% & 88.6\% \\
    \hline
  \end{tabular}
  \label{tab:accuracy}
\end{table}

\section{LIMITATIONS}

During the test week, 60 driving-over events where captured, resulting in a limited number of ground truth data. Especially for neural networks this number of events is quite small. Hence, the models may show overfitting.

Furthermore, the driving-over test where performed at low velocities. Hence, for higher velocities the signals may behave differently. 

 Objects that have not been driven over but have been pushed away continue to represent a challenge for the detection of drive-over events. Because the latter can also represent, for example, a collision with a living being, which is harder to detected due to the non-driving-over.

\section{CONCLUSIONS}

Driving-over detection in the railway environment is an important subsystem for fully automatic driving, since this task is currently done by the train driver. In a future railway system, this task has to be taken over by a automatic system, providing a similar or better performance. In this publication the first non-mechanically operating driving-over detection system for railway vehicles was proposed. 

For evaluating the approach, detailed experiments have been carried out. Based on the captured sensor data, three algorithms were developed for detecting driving-over events. 
The evaluation have shown that it is possible to detect driving-over events up to the maximal velocity tested, i.e. 15 km/h. Furthermore, the driving-over events can also be distinguish from regular events in rail operations, such as switches and skidding points. 

It could be determined which sensors modalities and the associated positions are suitable for the subsequent detection routine. Hereby, the acceleration sensors show the best results, while microphones, draw wire sensors and laser distance sensors where withdrawn.

For future work several open tasks have been identified.
In order to be able to establish a general validity of the results, further tests should take place especially in the higher velocity range.
The tests were done, with one type of test vehicle. Hence, further test vehicles should be included in tests, considering also operational maneuvers as shunting or pushing operations.

\begin{acknowledgments}
We would like to thank Havelländische Eisenbahn (HVLE) for providing the test track. Furthermore, we would like to thank Ben Noethlichs, Max Schischkoff, Thilo Hanisch, and Daniela Lauer for the support during the tests and the project. 

This work was supported by the Federal Ministry for Economic Affairs and Climat Action.
\end{acknowledgments}

\nocite{*}
\bibliography{IEEEexample}

\end{document}